\documentclass[%
reprint,
 amsmath,amssymb,
 aps,
]{revtex4-1}

\usepackage{graphicx}
\usepackage{dcolumn}
\usepackage{bm}
\usepackage{siunitx}

\begin{document}

\preprint{APS/123-QED}

\title{Measurement of Surface Acoustic Wave Resonances in Ferroelectric Domains by Microwave Microscopy}

\author{Scott R. Johnston}
\author{Yongliang Yang}
\author{Yong-Tao Cui}
\author{Eric Yue Ma}
\affiliation{Department of Applied Physics, Stanford University, Stanford, Ca 94305, USA}%

\author{Thomas K\"ampfe}
\author{Lukas M. Eng}
\affiliation{Institute of Applied Physics, TU Dresden, Dresden 01069, Germany}

\author{Jian Zhou}
\author{Yan-Feng Chen}
\author{Minghui Lu}
\affiliation{National Laboratory of Solid State Microstructures and Department of Materials Science and Engineering, Nanjing University, Nanjing, Jiangsu 210093, China}%

\author{Zhi-Xun Shen}
\affiliation{Department of Applied Physics, Stanford University}
\affiliation{Department of Physics, Stanford University, Stanford, Ca 94305, USA}

\date{\today}
\begin{abstract}
Surface Acoustic Wave (SAW) resonances were imaged within a closed domain in the ferroelectric LiTaO$_3$ via scanning Microwave Impedance Microscopy (MIM). The MIM probe is used for both SAW generation and measurement, allowing contact-less measurement within a mesoscopic structure. Measurements taken over a range of microwave frequencies are consistent with a constant acoustic velocity, demonstrating the acoustic nature of the measurement. 

\end{abstract}

\maketitle


\section{Introduction}

Ferroelectrics are materials with persistent electric polarization, similar to the persistent magnetization of ferromagnets. Ferroelectrics are piezoelectric, meaning that strain is strongly coupled to applied electric field. Because of this, they are used in a wide variety of applications, from strain sensors and actuators \cite{Sun1995} to the surface acoustic wave (SAW) based microwave delay lines \cite{Collins1968} and filters \cite{Tancrell1971} which are most relevant to this work.

SAW microwave devices such as SAW delay lines convert microwave electro-magnetic signals with centimeter wavelengths to acoustic waves confined to the surface of a crystal with micron wavelengths. This allows the design of miniaturized microwave delay lines, as well as high-performance filters and oscillators which take advantage of wave properties. While SAWs can propagate on the surface of non-piezoelectric materials \cite{Shilton1996}, piezoelectrics are uniquely suited to this application because the coupling between electric field and strain in a piezoelectric allows electrical signals to be converted directly into SAWs and back again.


This work focuses on the use of ferroelectric domain walls as SAW reflectors. The principle of acoustic reflection from ferroelectric domain walls has long been understood \cite{Kessenikh1971} and even used to create SAW delay lines with electronically-controlled variable delay \cite{Coldrern1977}. This reflection has been imaged strikingly in LiNbO$_3$ (LNO), a very similar ferroelectric to the LiTaO$_3$ (LTO) studied in this paper, by scanning electron microscopy \cite{Roshchupkin1994}.

Periodic ferroelectric domains have been used for both SAW generation \cite{Roshchupkin1994} and filtering \cite{Zhu1992}, with the wavelength set by the periodicity. More recently, reprogrammable, variable frequency SAW filters have been demonstrated by rewriting these periodic ferroelectric domain structures \cite{Ivry2014}.

All of these devices use large transducers to create surface acoustic waves, and additional transducers, or imaging techniques such as SEM, to observe them. 
In contrast, we have demonstrated a nano-scale method of generation and measurement of SAWs, allowing not only for SAW generation and measurements at high spatial resolution and large bandwidth, but also measurement of SAW behavior within a closed domain, such as the rounded triangular resonators described in this paper.

\section{Experiment and Theory}

\begin{figure}

\includegraphics[width = 7.5cm]{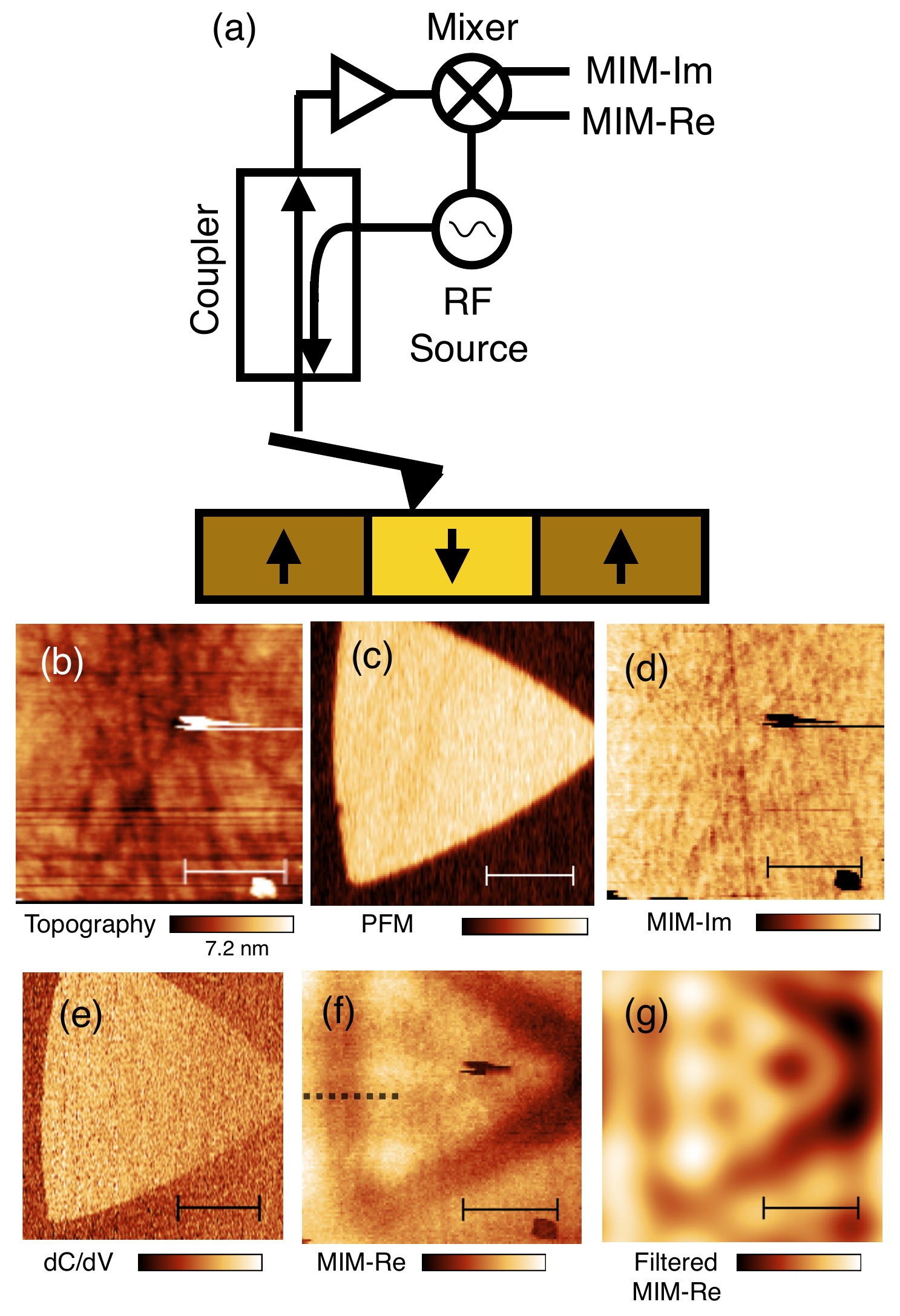}
\caption{(a) Schematic of the MIM measurement setup and sample. Topography, Piezo Force Microscopy (PFM) and MIM-Im images are shown in (b), (c), and (d) respectively. $dC/dV$, the change in MIM-Im with applied voltage, is shown in (e). MIM-Re is shown in (f), and in (g) is a 2D FFT filtered version of (f). All scale bars are 5 $\mu$m. Note: (c) and (e) were taken at different times using different tips.
}
\label{fig:overview}
\end{figure}

We examined periodically poled, polished, single crystal LiTaO$_3$ (LTO) using scanning Microwave Impedance Microscopy (MIM). The sample is a commercially available Z-cut crystal, with roughly \SI{15}{\micro m} polarized domains formed perpendicular to the polished surface as shown in Fig. \ref{fig:overview}. The ferroelectric domains in this sample were formed by applying kV potentials to periodic electrodes on the back of the sample while contacting the front of the sample by a conductive liquid. The front of the sample was then mechanically polished.

Measurements were done using our home-built MIM instrument \cite{Kundhikanjana2009} on a commercial atomic force microscope (AFM). The MIM electronics measure small changes in tip capacitance and dissipation of a custom, shielded AFM probe \cite{Yang2012a} as it is scanned across the dielectric sample, changing its self-capacitance. The basic principle is laid out in Fig. \ref{fig:overview}(a), and detailed in \cite{Lai2011}. Such measurements reach aF sensitivity levels and beyond \cite{Gramse2014}. This allows measurement at high spatial resolutions ($<40$nm), with spatial resolution set by tip size. Tips with radius of approximately 100nm were used for this study.

Piezo force microscpy (PFM) is a very common way to measure ferroelectric domain structure \cite{Abplanalp1998} in which a low frequency (\SI{}{kHz}) voltage is applied to the AFM tip and the piezoelectric response of the sample is measured using the the AFM's optical tip displacement detection. Because the sign of the small-signal piezoelectric response is dependent on the direction of polarization, the sign of the PFM response reveals the direction of polarization. Thus, the triangular domain shape is clearly shown in PFM (Fig. \ref{fig:overview} c). 

Ferroelectrics have a large dielectric constant which is strongly dependent on applied voltage (large $d\epsilon/dE$). The sign of this dependence is also determined by the direction of polarization. As a result, measuring the small-signal dependence of capacitance on applied voltage, $dC/dV$ provides clear domain contrast \cite{Cho2014}, as shown in figure \ref{fig:overview} e. 

The unmodulated capacative chanel, called MIM-Im because it is a measurement of imaginary admittance, only shows inverse topography, as shown in figure \ref{fig:overview} d. This is because the small increase in capacitance between the upper tip and the sample as the tip moves closer to the sample and the absence of any significant MIM-Im signal from material properties.

The loss channel, called MIM-Re for real admittance, displays clear features related to the domain (figure \ref{fig:overview} f), including a dark region microns wide at the domain wall and relatively faint patterns within the domain. These slowly varying patterns become more clear when the high-frequency components of the image (primarily noise and artifacts) are filtered out by discrete fourier transform filtering (figure \ref{fig:overview} g). 

We believe that the features in MIM-Re are the result of SAWs launched from the probe tip. The approximately \SI{1}{GHz} electric field at the tip causes the LTO to expand and contract due to the piezoelectric effect, creating SAWs at the tip, similar to other point source SAW generation methods such as focused laser pulses \cite{Maznev2003}. The microwave energy used to create mechanical motion is observed as loss (dissipation) in the MIM measurement. Thus, the MIM-Re image shows how much microwave power is coupled into acoustic waves at each position. 

If we assume that the tip is an acoustic point source, we expect to see less energy coupled into acoustic waves near the domain wall due to interference with reflected waves. Simplifying the experiment as a perfect point source in an isotropic plane, this dip in loss near the domain wall should be approximately half a SAW wavelength in width ($\lambda/4$ on each side), following the equation:
\begin{equation}
\label{eq:dissx}
D(x) = D_0 \times (1 - J_0(2kx))
\end{equation}
where $k = 2\pi/\lambda $, $J_0$ is the 0-order J-type Bessel function, and $D_0$ is an arbitrary factor. This equation is derived in the Appendix \ref{app:calc}.

The wavelength, and consequently the width of the dissipation trough at the domain wall, should vary inversely with the excitation frequency ($\lambda f = v$). To test this, MIM measurements of a domain wall were taken at a variety of frequencies between \SI{700}{MHz} and \SI{1300}{MHz} and the spatially resolved dissipation around the domain wall compared between frequencies.

\section{Results and Discussion}

\begin{figure}

\includegraphics[width = 7.5cm]{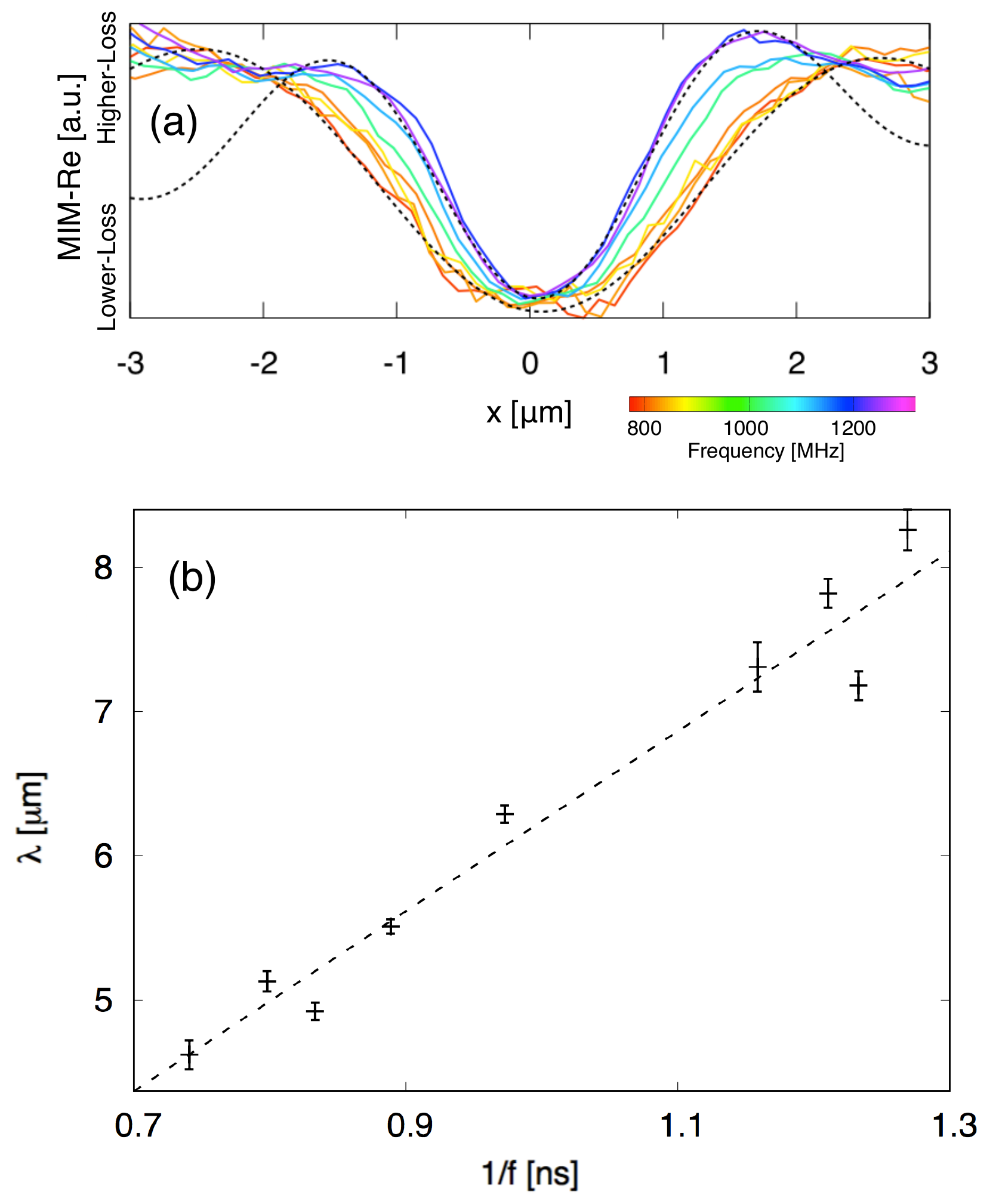}
\caption{(a) A series of normalized line-cuts taken at the same location with different measurement frequencies. The location of the line-cuts is indicated in Fig. \ref{fig:overview} f by a dashed line. Each line-cut has been fitted by equation \ref{eq:dissx}. Fits for the highest and lowest frequency lines are shown with dashed lines. (b) Wavelength at each frequency as determined by the fits in (a). A fit for constant velocity (linear with no offset) is shown with a dashed line and gives propagation velocity of \SI{6.24 \pm .09}{km/s} (fitted line). Error bars are derived solely from the fits in (a), and do not account for all sources of error ($\chi_r^2 = 10.0$).
}
\label{fig:halo}
\end{figure}

The dip in loss around the domain wall decreases in width as microwave frequency is increased, as is clearly visible in Fig. \ref{fig:halo} a. It is well fitted by equation \ref{eq:dissx} within the trough, though the tails differ significantly because of additional complexities of the system, principally that the domain is triangular rather than linear, as well as the significant anisotropy of the actual crystal. Note that for this reason the errors on $\lambda$ derived from each fit underestimate the true error, which is due to these systematic errors in addition to noise.
The value of $\lambda$ from each fit is shown in Fig. \ref{fig:halo} b. We then fit these points by a line of constant velocity, extracting a SAW velocity of \SI{6.24 \pm .09}{km/s}. 

\begin{figure}

\includegraphics[width = 7.5cm]{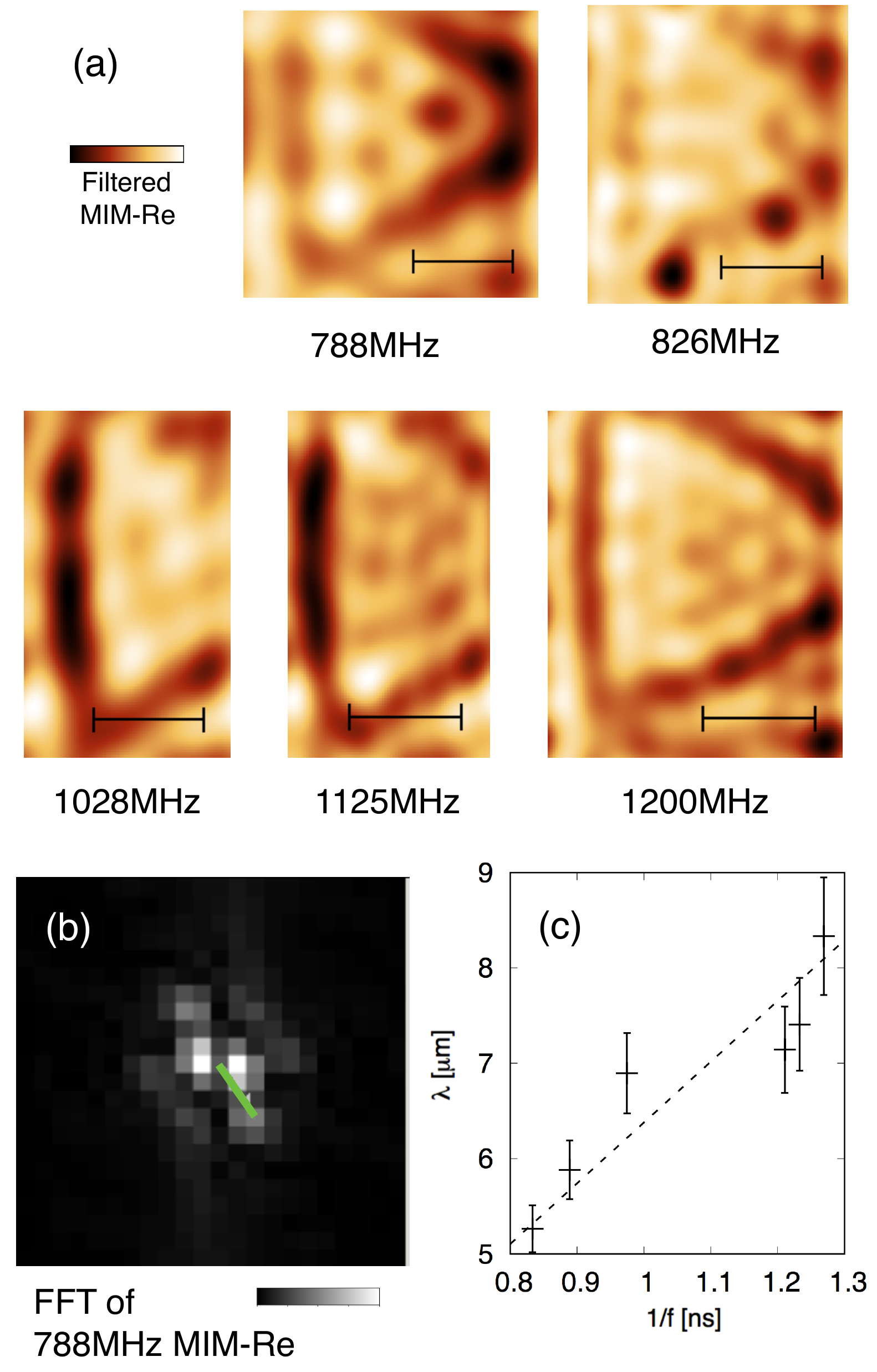}
\caption{(a) 2D FFT Filtered MIM-Re images showing the change in SAW modes with frequency. (b) Example of Fourier transformed image from (a), showing a ring of peaks with distance to one indicated in green. (c) Distance to the same Fourier peaks at different frequencies, fitted by a line of constant velocity. This fit gives a SAW velocity of \SI{6.38 \pm .16}{km/s}, in good agreement with the fit in FIG. 2. c. Errors are given by the discrete Fourier pixel size and account for most of the observed error ($\chi_r^2 = 1.2$).
}
\label{fig:FFT}
\end{figure}

In addition to the frequency dependent width of the decreased dissipation near the domain walls, we also observe frequency dependent patterns within the domain. These decrease in spacing as wavelength is decreased, as pictured in Fig. \ref{fig:FFT} a. We expect the spots of high (and low) loss to be spaced at approximately $\lambda/2$, consistent with the periodicity of equation \ref{eq:dissx}. This correspondence is not necessarily exact because of the discrete modes of the closed cavity, but will still hold on average.

To show this, we've taken 2-dimensional discrete Fourier transforms of the data and measured the distance to the peaks shown in Fig. \ref{fig:FFT} b. These peaks correspond to the inverse spacing of the features observed in Fig. \ref{fig:FFT} a. Plotting twice the distances from the Fourier transforms versus inverse frequency, we observe agreement with a zero-offset linear fit (constant velocity). Furthermore, the fit gives a SAW velocity of \SI{6.36 \pm .16}{km/s}, in good agreement with the measurement from the width of the trough at the domain wall of \SI{6.24 \pm .09}{km/s}. While we expect errors from the discrete cavity modes, most of the error observed is due to the pixel size of the discrete Fourier transform ($\chi_r^2 = 1.2$)



It is surprising that the velocities we measure are not consistent with the velocity of the Raleigh SAW mode typically seen in SAW devices, which has a velocity around 3.2 to \SI{3.4}{km/s} \cite{Slobodnik1972}. One possibility is that tip pressure and peculiarities of our geometry cause the unexpected dominance of the longitudinal leaky SAW mode, which has a velocity around \SI{6}{km/s} in other cuts of LTO \cite{Tonami1995}\cite{Saw1999}, though it is expected to have high loss on the Z-cut surface \cite{Kenny2011}, and is consequently not well studied on this surface.

It is also possible that what we see is the result of a slower SAW mode, and a more nuanced analysis would explain how our results arise from a slower SAW velocity. Further study of the details of SAW generation by AFM tips and interference within a closed domain is likely needed for a complete understanding of the measurement.

\section{Conclusions and Outlook}

We have shown that MIM can be used to measure acoustic properties of piezoelectric devices. Going forwards, we hope this technique can be applied to probe various SAW devices, such as the SAW resonators which are currently of interest for their coupling to superconducting qubits \cite{Manenti2016}. It is also possible that this technique could be used to rapidly pattern and probe reprogrammable SAW devices which make use of ferroelectric domain walls \cite{Ivry2014}.


In addition, it is important that SAW effects near domain walls be accounted for in microwave studies of ferrolectric domain wall conduction.
Specifically, in studies done at higher frequencies (\SI{3}{GHz}) with a dull tip (100s of nm) \cite{Tselev2016}, the dip in MIM-Re at the domain wall could be impossible to distinguish from the increase due to domain wall conduction, leading to an erroneously low measurement of the AC conductivity. This discrepancy can be accounted for by taking measurements with resolution well below the SAW wavelength, allowing the differentiation of the short range signal from the thin, conductive domain wall from the longer length-scale variations due to SAWs. This understanding is particularly relevant for LTO, which has been shown to host conductive domain walls under certain conditions \cite{Cho2014}.

In summary, we have demonstrated the ability to measure SAW properties in closed resonators in the ferroelectric LiTaO$_3$ using scanning microwave impedance microscopy over a range of frequencies.

\section*{Appendix}

See Appendix \ref{app:calc} for a derivation of equation \ref{eq:dissx}.

\section*{Acknowledgements}

The authors would like to acknowledge the support of NSF grant DMR1305731, The Gordon and Betty Moore Foundation, EPiOs grant GBMF4536, and The National Nature Science Foundation of China, grant 11134006. We would also like to thank the creators and contributors of the free, open source software used for data analysis and plotting, gnuplot and gwyddion, and helpful conversations with Keji Lai.

\section*{Author Contributions}
S.R.J., Y.Y., and T.K. performed measurements. S.R.J and Y-T.C. analyzed and interpreted the data. J.Z., M.L., and Y-F.C. fabricated the sample. All authors discussed the results. S.R.J. wrote the paper. Z-X.S. oversaw the project.

\bibliography{library}

\clearpage

\appendix
\section{\label{app:calc} Energy coupling of point source next to reflecting boundary}

To calculate the energy coupled into surface acoustic waves near a domain wall, we model the tip as a point source wave generator and the surface of the ferroelectric as a two-dimension plane which carries these waves. The domain wall is modeled as an infinite reflector. Additional complexities such as anisotropy is wave velocity are neglected. To calculate the average power coupled in, we calculate the power radiated in the plane as a function of the distance between the tip and the domain wall. To do this, we'll look at the power passing through a circle containing the tip and far away relative to the tip domain-wall distance.

Let $x_0$ be the distance between the tip and the domain wall, with the domain wall positioned at $x=0$ and the tip at $x_0$. We can satisfy the boundary condition of a perfectly reflecting domain wall (displacement, $u(0,y) = 0$) by placing a mirror source with opposite phase at $x = -x_0$. Thus, total displacement is given by:
\begin{equation}
\label{eq:a1}
\begin{split}
u(\vec{r},t) = \frac{A_0}{\sqrt{\lvert \vec{r} - \vec{x}_0 \rvert}}\sin(k\vert \vec{r} - \vec{x}_0\vert - \omega t + \phi_0) \\ 
+ \frac{A_0}{\sqrt{\lvert \vec{r} + \vec{x}_0 \rvert}}\sin(k\vert \vec{r} + \vec{x}_0\vert - \omega t + \phi_0 + \pi)
\end{split}
\end{equation}
where $A_0$ is arbitrary. In polar coordinates, $\vert \vec{r} \pm \vec x_0 \vert = \sqrt{r^2 + x_0^2 \pm 2 r x_0 \cos(\theta)}$. We can Taylor expand this to $r\left(1 \pm \frac{x_0}{r}\cos{\theta} + \mathcal{O}((\frac{x_0}{r})^2)\right)$ and we are only interested in $r>>x_0$.  

Since the zero order terms in $\frac{x_0}{r}$ of $\frac{A_0}{\vert \vec{r} - \vec{x}_0 \vert}$ do not cancel out of equation \ref{eq:a1}, we can rewrite it as:
\begin{equation}
\begin{split}
u(\vec{r},t) = \frac{A_0}{\sqrt{r}}[\sin(k\vert \vec{r} - \vec{x}_0\vert - \omega t + \phi_0) \\ 
+ \sin(k\vert \vec{r} + \vec{x}_0\vert - \omega t + \phi_0 + \pi)]
\end{split}
\end{equation}
Converting the sum of sines to a product we get time dependent and independent parts:

\begin{equation}
\begin{split}
\label{eq:timesplit}
u(\vec{r},t) = \frac{2A_0}{\sqrt{r}}\sin\left(\frac{k}{2}(\vert \vec{r} - \vec{x}_0\vert - \vert \vec{r} + \vec{x}_0\vert)\right) \\
\times \cos\left(\frac{k}{2}(\vert \vec{r} - \vec{x}_0\vert + \vert \vec{r} + \vec{x}_0\vert) +\pi_0 + \pi - \omega t\right)
\end{split}
\end{equation}

We are only interested in the power carried away by the wave, so we can ignore the time dependent portion and focus only on the time-independent envelope (top half of equation \ref{eq:timesplit}), which can be written to first order in $\frac{x_0}{r}$ as:

\begin{equation}
A(\vec{r}) = \frac{2A_0}{\sqrt{r}}\sin(k x_0 \cos\theta)
\end{equation}
\\
The power is proportional to $A^2(r)$, with the total power crossing a circle of raidus $r$ given by:
\begin{equation}
\begin{split}
\int_0^{\pi}A^2(r,\theta)\frac{1}{r}d\theta = 4A_0^2 \int_0^{\pi} \sin^2(kx_0\cos\theta)\\ \propto 1 - J_0(2kx_0)
\end{split}
\end{equation}

where $J_0$ is the 0-order J-type Bessel function.

\end{document}